\documentclass[12pt]{article}

\newcommand{\bu}{\Upsilon}
\newcommand{\epb}{\mbox{b}}
\newcommand{\epba}{\overline{\mbox{b}}}
\newcommand{\epc}{\mbox{c}}
\newcommand{\epca}{\overline{\mbox{c}}}

\newcommand{\epe}{\mbox{e}^-}
\newcommand{\epea}{\mbox{e}^+}
\newcommand{\epQ}{\mbox{Q}}
\newcommand{\epq}{\mbox{q}}
\newcommand{\epQa}{\overline{\mbox{Q}}}
\newcommand{\epqa}{\overline{\mbox{q}}}
\newcommand{\eps}{\mbox{s}}
\newcommand{\epsa}{\overline{\mbox{s}}}

\newcommand{\ggn}{\mbox{g}}
\newcommand{\gwm}{\mbox{W}^-}
\newcommand{\gwp}{\mbox{W}^+}

\newcommand{\bibl}[5]
	{#1, {\it #2} {\bf #3} #5 (#4)}
\newcommand{\cebe}{\begin{center}}
\newcommand{\ceen}{\end{center}}
\newcommand{\debe}{\begin{description} \vspace{-2ex}}
\newcommand{\deen}{\end{description}}
\newcommand{\eabe}{\begin{eqnarray}}
\newcommand{\eaen}{\end{eqnarray}}
\newcommand{\epem}{$\mbox{e}^+\mbox{e}^-$}
\newcommand{\eqbe}{\begin{equation}}
\newcommand{\eqen}{\end{equation}}

\newcommand{\ra}{\rightarrow}
\newcommand{\tr}{\mbox{Tr}\;}

\input{psfig}

\newcommand{\Kbar}{\overline{\mbox{K}}}
\newcommand{\tm}{\mbox{T}}

\begin{document}

\begin{titlepage}
\begin{flushright}
  LU TP 96-10 \\
  April 1996
\end{flushright}
\vspace{25mm}
\begin{center}
  \Large
  {\bf Colour connections in \epem-annihilation} \\ 
  \normalsize
  \vspace{12mm}
  Christer Friberg, G\"osta Gustafson, Jari H\"akkinen\footnote{christer@thep.lu.se, gosta@thep.lu.se, jari@thep.lu.se} \vspace{1ex} \\
  Department of Theoretical Physics, Lund University, \\
  S\"olvegatan 14A, S-223 62 Lund, Sweden
\end{center}
\vspace{20mm}

\noindent {\bf Abstract:} \\
We have a very limited knowledge about how the confinement mechanism works in processes like  $\epea\epe\ra\epq\epqa\ggn\ldots\ggn$, when there are identical colour charges. In this case the partons can be connected by a string or a cluster chain in several different ways. We do not know if in such a situation Nature chooses a particular configuration at random, or if some configuration is dynamically favoured. Also in the perturbative parton cascade we have no well founded recipe for describing the interference effects, which correspond to non planar diagrams and are caused by identical colour charges. We have studied two different models, and are in particular interested in the possibility that a colour singlet gluon system hadronizes isolated from the remainder of the state. Using double tagged events with heavy $\epc\epca$ or $\epb\epba$ quarks, it appears to be possible to find a significant signal, if such events appear in Nature.

If this type of (re)connected states appear in Z decays it may also be an indication that reconnection might appear between the decay products of two W's in the reaction $\epea\epe\ra\gwp\gwm\ra\epq_1\epqa_2\epQ_1\epQa_2$ at LEP2. This would be important e.g. for a precision measurement of the W mass.
\end{titlepage}

\section{Introduction}
High energy reactions, e.g. $\epea \epe$-annihilation, are generally described in terms of two phases: First a short time perturbative phase with large momentum transfers, described in terms of quarks and gluons. This initial phase is described by a parton cascade. Secondly a long time soft interaction phase, in which the energy of the partons is transformed into hadrons. This non-perturbative hadronization phase is successfully described by a string model~\cite{ba83} or a cluster model~\cite{bw84}. These two phases are generally assumed to be well separated from each other. The parton cascade has a cutoff of the order 0.5--1GeV corresponding to a timescale of 0.2--0.4fm, while the hadronization time often is assumed to be 1--2fm. It is believed that the reaction cross section is fully determined by the perturbative phase, while in the soft hadronization phase a definite hadron state is chosen with total probability 1.

In both phases a set of approximations is used, where the number of colours is assumed to be large. Assuming infinitely many colours reduces the possible interference effects. Only planar diagrams contribute in the perturbative cascade, and the way to connect the partons by a string or cluster chain becomes unique. A specific colour charge has to be connected to a parton with the corresponding anti-colour, and with infinitely many colours the probability that two (or more) partons have the same colour is zero.

With only three colours, as in Nature, this is no longer the case. The basic problem is that the direct correspondence between the parton states and the string states (or cluster chains) is lost. The parton state is determined by the momenta, polarization and colour of all the quarks and gluons. This is not enough to describe the true states in QCD. Also some topological quantum numbers have to be specified, related to the vacuum condensate and the confinement mechanism. 't~Hooft has proposed that these extra quantum numbers correspond to gauge singularities similar to the ones appearing in a superconductor~\cite{gh81}. In the string model they correspond to the string topology; the string state is specified by the momenta carried by the string ends (quarks and anti-quarks) and by the corners (gluons), together with a specification of how they are connected.

\begin{figure}[t,b]
  \hbox{\vbox{
    \begin{center}
    \mbox{\psfig{figure=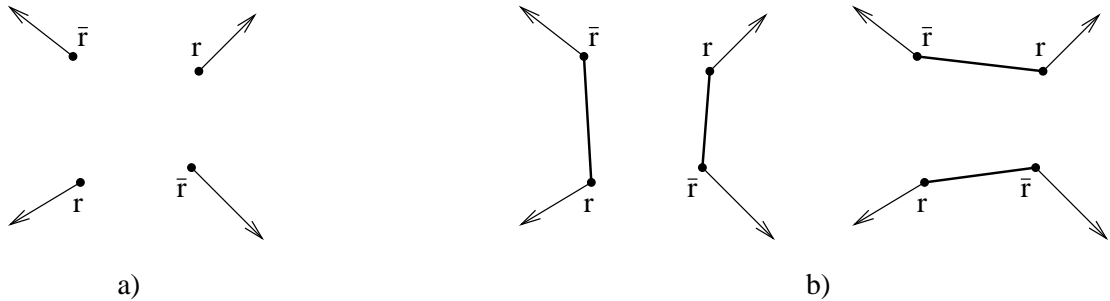,height=4cm}}
    \end{center}
  }}
\caption{\em When two red and two anti-red colour charges move apart they can be connected by strings or cluster chains in two ways.}
\label{f:4quark}
\end{figure}

The most simple example is shown in Fig~\ref{f:4quark}. With two red quarks and two anti-red anti-quarks there are two different possible string states. The string configuration is in principle an observable (reflected in the momenta of the final state hadrons), which corresponds to excitations in the vacuum condensate. When the coloured partons move apart, the confinement mechanism must give a response in the condensate, such that a particular string state is selected. This selection cannot be determined from perturbative QCD, because the distinction between the two string states in Fig~\ref{f:4quark}b does not correspond to any observable related to the parton state in Fig~\ref{f:4quark}a.

The fact that the present models work so well, indicates that these effects are usually small. They are generally of the order $1/N_c^2$, which corresponds to the probability that two gluons have the same colour. They are nevertheless of fundamental importance for our understanding of the confinement mechanism. What determines the response of vacuum to the produced partons, and what is the time scale of the ``interface'' between the hard and the soft phases? When identical parton charges allow several state topologies, is one topology chosen at random or is some configuration dynamically favoured? Connected to these problems are colour suppressed contributions to the perturbative parton cascade, related to non planar diagrams.

As the problems cannot be solved by perturbative QCD we have to rely on experimental measurements. Thus in this paper we want to study possible effects of the finite number of colours, and observables which are sensitive to these effects. The problem has been studied in refs~\cite{gg94,gg88,ts94:2,ll95} in connection with the reaction $\epea \epe \ra \gwp \gwm \ra \epq_1 \epqa_2 \epq_3 \epqa_4$\footnote{In the calculations in~\cite{gg94} the polarization of the W's was neglected. When this is taken into account the reported signal is reduced by approximately a factor 1/2}. At LEP2 the two $\epq \epqa$ pairs originate from points separated by about 0.1fm, which is short compared to the hadronization scale. Here we will study similar effects in the reaction $\epea \epe \ra \epq \epqa$ + a number of gluons at LEP1. With few colours it is possible to obtain a gluonic subsystem, which is in a colour singlet state. This system could hadronize independently, decoupled from the remaining state (in the string model as a closed string), and we want in particular to study if it is possible to experimentally identify such gluonic colour singlets. At LEP1 the whole process starts in a single point, and a comparison of the effects in the two reactions would be very valuable.

There are three different problems we want to address in this paper: \vspace{-2ex}
\begin{itemize}
\item What can we learn about the interface between the hard and the soft phases? \vspace{-1ex}
\item How does the confinement mechanism work when there are several ways to connect the partons by a string? \vspace{-1ex}
\item Is it possible to find a good approximation scheme for the interference effects in the perturbative phase? \vspace{-1ex}
\end{itemize}
These problems will be further discussed in the following sections. The last problem is in principle solvable within perturbative QCD, but this is not the case for the other two problems. We will describe two different models and discuss how these can be experimentally tested. The effects on average event properties are generally small, but it is possible to find an observable signal for special events.

\section{Interface between the hard and soft phases}
\begin{figure}[t,b]
  \hbox{\vbox{
    \begin{center}
    \mbox{\psfig{figure=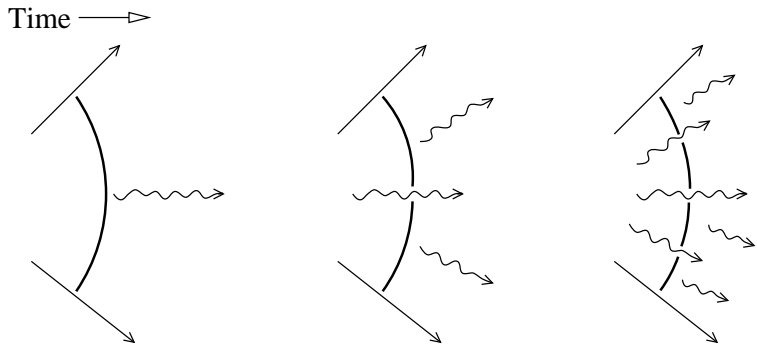,height=4.5cm}}
    \end{center}
  }}
\caption{\em The dipole picture of a gluon cascade.}
\label{f:cascade}
\end{figure}

In an $\epea \epe$-annihilation event full perturbative calculations have been performed only to second order in $\alpha_s$. This is far from sufficient to reproduce experimental data, and different approximation schemes have been developed to approximate the evolution of a parton cascade. In the dipole cascade model~\cite{gg86}, an initial colour dipole between a quark and an anti-quark can emit a gluon and thereby split in two dipoles, which can emit further gluons in a cascade as illustrated in Fig~\ref{f:cascade}. The dipoles form a chain where the colour of one gluon is combined with the anti-colour of a neighbouring gluon. The colour and the corresponding anti-colour radiate coherently producing the 'dipole emission'. This effect is often called soft gluon coherence~\cite{am81}, and can be approximated by an angular ordering of the emitted gluons. In the large $N_c$ limit all the dipoles have different colours, and therefore radiate independently, apart from the recoils experienced due to the emission. With only three different colours we must frequently have a situation with e.g. two red and two anti-red charges, which have to interfere in the further gluon emission. In the original dipole cascade model, implemented in the Ariadne MC~\cite{ll92}, this interference is neglected, and also in other versions of the parton cascade, e.g. those implemented in the Herwig~\cite{gm92} or the Jetset~\cite{ts94} MC, the only interference included is the one from a colour charge and the directly associated anti-charge. One approach to include interference effects, which we will discuss further below, is implemented in the latest version of Ariadne (version 4.07).

\begin{figure}[t,b]
  \hbox{\vbox{
    \begin{center}
    \mbox{\psfig{figure=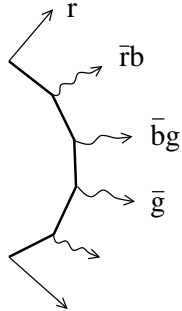,height=4.1cm}}
    \end{center}
  }}
\caption{\em In the string model a string-like colour field is stretched from red to anti-red, from blue to anti-blue etc.}
\label{f:simpletopology}
\end{figure}

Thus in the perturbative cascade a quark-gluon state with well specified colours is produced. Via the confinement mechanism the energy of these partons is transformed into colour singlet hadrons. In the string model a string-like force field is stretched from a quark via a set of gluons to an anti-quark. Thus in the example in Fig~\ref{f:simpletopology} the string is stretched from red to anti-red, then from blue to anti-blue etc.. With many colours the string configuration corresponds exactly to the chain of dipoles, but if there are more than one red--anti-red pair there may be more than one possible way to stretch the string. 

As mentioned in the introduction the string configuration appears as an observable. Even if a certain string state does not correspond uniquely to a definite hadron state, different possible string states produce usually very different hadronic states, and they can therefore be distinguished from each other. We note that in the string model also the direction of a closed string, corresponding to a purely gluonic system, is an observable (with the direction defined e.g. from colour to anti-colour). When the string breaks by the production of an $\eps \epsa$ pair the s and $\epsa$ are pulled in opposite directions, which implies a rapidity separation between e.g. a final state K and $\Kbar$. This means that e.g. the decay $\Upsilon \ra$ 3g can give two differently oriented string loops, which cannot be separated by perturbation theory~\cite{gg82}.

Thus the string or cluster chain configuration is an observable, which is not fully determined by the partonic state of coloured quarks and gluons, and the confinement mechanism has to imply that a certain configuration is selected. At which stage in the process is this configuration fixed? Is this fixation purely in the hadronization phase, or does it also affect the initial hard phase. To clarify this problem we will compare with the behaviour of a superconductor, and discuss in more detail the example of $\Upsilon$ decay.

If the vacuum condensate behaves like a type I superconductor, we expect that the bag model is a good representation of the confinement mechanism, and the string should be more regarded as a flux tube with a radius of the order of 1fm. It is then natural to expect that the topology of the string or cluster chain is not determined until the partons are separated by a similar distance, and that this process is totally separated from the perturbative cascade.

If on the other hand vacuum is more like a type II superconductor the string would be more similar to a vortex line, where the energy is concentrated to a thin core, although the field can be smeared out over a larger region. We can then imagine that the topology is fixed at an earlier stage when the parton separation is of the order of the core radius. In this case the mechanism which determines the string configuration could possibly also affect the early hard phase of the reaction.

\begin{figure}[t,b]
  \hbox{\vbox{
    \begin{center}
    \mbox{\psfig{figure=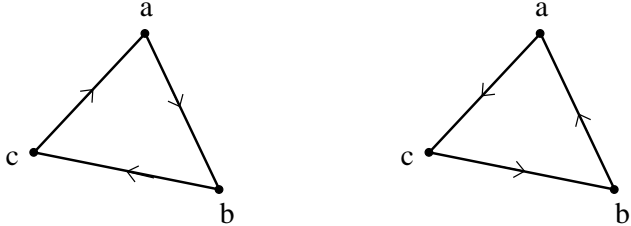,height=3.0cm}}
    \end{center}
  }}
\caption{\em Two different string configurations are possible in the decay $\bu\ra 3\ggn$. The direction of the string is defined as going from colour to anti-colour.}
\label{f:orient}
\end{figure}

As an example let us study the reaction $\Upsilon \ra$ 3g. The decay amplitude is proportional to the colour factor
\eqbe
\frac{1}{2} d^{abc} = \tr [ \tm^a \tm^b \tm^c ] + \tr [ \tm^c \tm^b \tm^a ] \equiv A + B
\eqen
where $a$, $b$, and $c$ are the colours of the three gluons. The two terms correspond to different orientations of the string (see Fig~\ref{f:orient}).
In the large $N_c$ limit the two amplitudes are orthogonal, and the cross section for the two different string states is given by the colour factors (obtained by summing over all colours)
\eqbe
\sum |A|^2 = \sum |B|^2 = \frac{(N_c^2 - 1)(N_c^2 - 2)}{8 N_c^2}
\eqen
For finite $N_c$ there is however also an interference term
\eqbe
\sum AB^\dagger = - \frac{N_c^2 - 1}{4 N_c} 
\eqen
which corresponds to a part of the cross section, for which the orientation of the final state is not determined. An example is when all three gluons are $r \overline{r}$. Although formally of order $1/N_c^2$ this part is clearly not small; it is negative and for three colours it is a fraction $2/(N_c^2 - 4)$ = 40\% of the total cross section. Since the two states in Fig~\ref{f:orient} are eigenstates to an observable with different eigenvalues, the full amplitudes have to be orthogonal. If the topology is fixed at an early stage of the process, it could conceivably interfere with the perturbative phase and thus influence the cross section~\cite{gg82}. If this would imply that the interference term should be neglected in a calculation of the cross section, a determination of $\alpha_s$ to fit experimental data would give a smaller value, not in agreement with $\alpha_s$ measurements at higher energies~\cite{alpha}. Although not a proof, this is at least an indication that the string configuration is determined at a later stage well separated from the initial hard process, which determines the cross section.

Thus, if the string topology is not determined in the hard phase we must face the following problem: Sometimes the initial perturbative process produces a state with several identical colour charges moving apart. What mechanism selects a specific final state topology? Is the topology chosen at random among those which are possible, or is some topology dynamically favoured? In the next section we will discuss some conceivable models.

\section{Models}
\subsection{The hadronization phase}
\label{ss:hadron}
\begin{figure}[t,b]
  \hbox{\vbox{
    \begin{center}
    \mbox{\psfig{figure=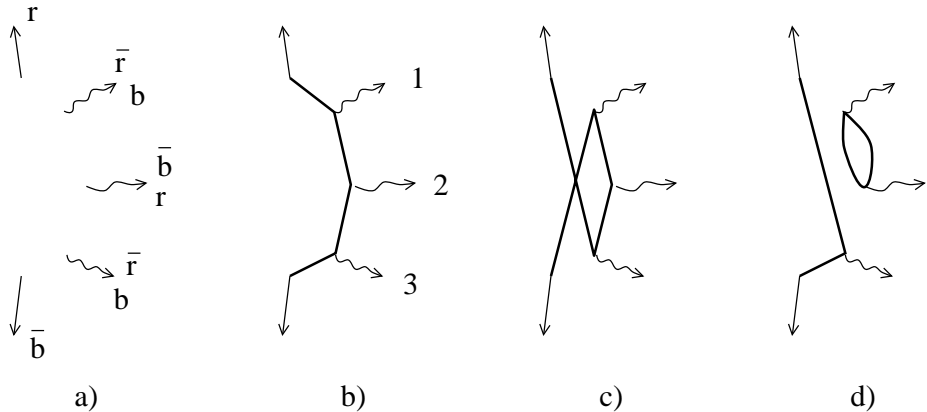,height=5.4cm}}
    \end{center}
  }}
\caption{\em The partons in a) can be connected in three different ways as in b), c), and d).}
\label{f:topology}
\end{figure}

We first want to discuss a model (called the soft reconnection model) in which the perturbative cascade is unchanged, and only the hadronization phase is modified. Let us assume that the perturbative phase has resulted in a set of partons with colour charges as shown in Fig~\ref{f:topology}a. These charges can be connected by a string (or cluster chain) in different ways as illustrated in Fig~\ref{f:topology}b--d. As discussed above we do not know if Nature chooses one configuration at random, or if some particular configuration is dynamically favoured. The same problem is studied in~\cite{gg94,gg88,ts94:2,ll95} in connection with the reaction $\epea \epe \ra \gwp \gwm \ra \epq_1 \epqa_2 \epq_3 \epqa_4$. As in ref~\cite{gg94} we will here study a model where the confinement mechanism preferentially connects a colour with a corresponding anti-colour in the neighbourhood. This implies that ``short'' string configurations should be favoured, and to decide what is a ``short string'' we will, as in~\cite{gg94}, use the $\lambda$ measure defined in~\cite{ba88}. This measure corresponds to an ``effective rapidity length'' , and is closely correlated to the hadronic multiplicity. For hard gluons it is approximately given by $\lambda \approx \sum \ln \left[(p_i + p_{i+1})^2/m_0^2\right]$ where $p_i$ are the (ordered) parton momenta and the hadronic mass scale $m_0$ is around 1GeV.

Assume that the ordering in Fig~\ref{f:topology}b is the one generated in the hard phase. Then the configuration in Fig~\ref{f:topology}c is possible provided the gluons numbered 1 and 3 have identical colours. Similarly the configuration in Fig~\ref{f:topology}d is possible if the colour of gluon 1 corresponds to the anti-colour of gluon 2. As there are 8 different gluon colours, these possibilities appear in general with probability 1/8. The SU(3) colour algebra will be discussed further in Section~\ref{s:cstructures}, but the basic property of our model is that a possible colour recoupling of the partons, which results in a smaller value for the $\lambda$ measure, will be chosen with a probability 1/8. (One constraint is that two or more gluons, which form an isolated colour singlet system as in Fig~\ref{f:topology}d, must not originate from a single colour octet gluon.)

The possibility to connect the partons as in Fig~\ref{f:topology}c or ~\ref{f:topology}d is often called colour reconnection or recoupling. This is actually an improper terminology. The partons have not been {\em re}connected. They did not start connected as in Fig~\ref{f:topology}b, and colour connection would be a better term. This is similar to the use of the term recombination time for the time when electrons and protons first combine to neutral atoms after the Big Bang.

\subsection{The hard phase}
\label{ss:hardphase}
\begin{figure}[t,b]
  \hbox{\vbox{
    \begin{center}
    \mbox{\psfig{figure=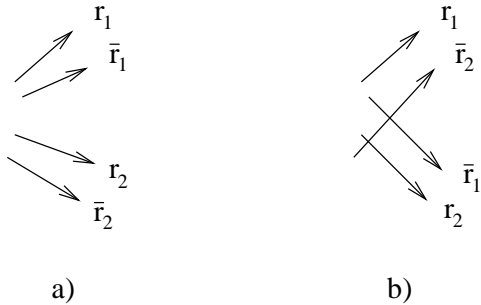,height=4.0cm}}
    \end{center}
  }}
\caption{\em {\em a)} Usually a colour charge and its associated anti-colour are close in phase space. {\em b)} Occasionally a red charge can be closer to an anti-red charge from a different pair.}
\label{f:4colour}
\end{figure}

In a second model also the hard perturbative phase is modified. In principle this phase is calculable within perturbative QCD. This is not possible in practise, and as mentioned above, in most parton cascade models all non-planar diagrams are neglected. (An exception is the latest version of Ariadne, which we will comment on further below.) This means that essentially only interference effects from e.g. a red and its adjacent anti-red charge in the same dipole is included. Assume that two r$\overline{\mbox{r}}$ pairs are produced. In principle they would emit further soft gluons as a quadrupole. In most cases the two partners in one dipole are close in phase space, and therefore appears as a neutral object when viewed from the other dipole, as illustrated in Fig~\ref{f:4colour}a. In these cases the neglected interference between the dipoles should be small. Occasionally this is not true, and the situation can look as in Fig~\ref{f:4colour}b. Here a better approximation would be to include interference within the pairs $r_1 \overline{r}_2$ and $r_2 \overline{r}_1$, but neglect the interference between these pairs. This means that the emission of softer gluons can correspond to a different ordering of the charges in the dipole chain.

Since a proper inclusion of the interference effects is too complicated, we want to study a rather simple model, assuming that this model can give information about possible qualitative features and the expected order of magnitude of the effects. In the parton cascade it can happen that the colour charges can be arranged in a dipole chain in different ways. In the model we assume that in such cases the emission of softer gluons is better approximated if we reorder the gluons, so that the charges in the dipoles are as close in phase space as possible. As a measure, when we compare two different orderings, we also here use the $\lambda$ measure discussed above.

As in the soft reconnection model the possibility to have identical gluon colours is taken into account by allowing the reconnection with probability 1/8, if it provides a smaller $\lambda$ measure.

Thus, at every step in the cascade we check if the gluon chain should be reordered. The colour reconnection of the gluons should be relevant for the emission of softer gluons later in the cascade. The Landau-Pomeranchuk formation time in a gluon emission is related to its transverse momentum, and therefore it is essential that  the transverse momentum is used as an ordering parameter in the cascade. For this reason the dipole cascade model, implemented in Ariadne, is particularly convenient, and we will use a modification of this program, in spite of the problems discussed below.

A model in which reconnection can occur {\em both} in the hard phase and in the soft phase will in the following be called the cascade reconnection model.

\section{Colour structures and mother-daughter relations}
\label{s:cstructures}
\subsection{Colour structures}
\begin{figure}[t,b]
  \hbox{\vbox{
    \begin{center}
    \mbox{\psfig{figure=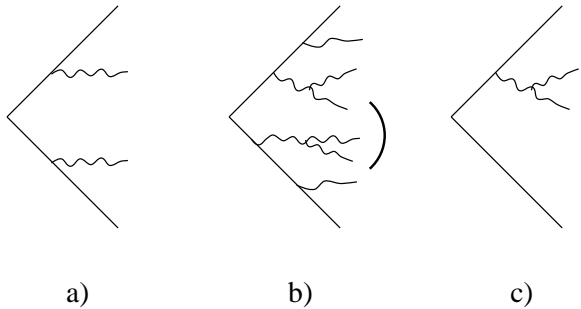,height=4.1cm}}
    \end{center}
  }}
\caption{\em {\em a)} A pair of gluons emitted from (anti-)quark legs are in a colour singlet state with probability $1/\left[N^2_c-1\right]=1/8$. {\em b)} This is also approximately the case for any connected gluon sub-system. {\em c)} An exception is two or more gluons originating from a single gluon parent. Such a system is naturally always a colour octet.}
\label{f:cstructure}
\end{figure}

We are particularly interested in the presence of isolated gluonic systems, as in Fig~\ref{f:topology}d, because such states can possibly be experimentally identified. Such an isolated system must clearly form a colour singlet state. A gluon pair produced as in Fig~\ref{f:cstructure}a is a colour singlet with probability $1/(N^2_c-1)$. This is also approximately the case for any connected gluon system, as e.g. the one in Fig~\ref{f:cstructure}b. (We note that with 5 gluons, which is the average number in Ariadne at the $Z^0$-pole, we have 10 different connected gluon systems. If all of these would have the probability 1/8 to form an isolated gluonic system, such systems would not be rare.)

An exception is two or more gluons originating from a single initial gluon as in Fig~\ref{f:cstructure}c. Such a system is always a colour octet. Thus it is essential to know if a gluon is emitted by a parent quark or a parent gluon. This is no problem in the cascade schemes implemented in the Herwig or the Jetset MC's. However, in the dipole cascade model a gluon is not emitted from a single parent, but is the result of the separation of a colour charge and anti-charge in a colour dipole. Furthermore colour coherence implies that a gluon emitted from a dipole between two parent gluons can be effectively emitted from a parent quark (i.e. colour triplet) charge. 

This problem was studied in~\cite{gg93} and some results will be discussed in the next subsection. The colour factor for gluon emission from a quark is $C_F=\frac{1}{2}N_c\left[1-1/N_c^2\right]$ , which is slightly less than half the corresponding factor for emission from a gluon (which is connected to two dipoles), $C_A=N_c$. The aim of ref~\cite{gg93} was to study how to take into account this colour suppressed contribution to the cascade evolution, but the results can be used also for our purpose analysing the mother-daughter relation.

\subsection{Mother-daughter relations in the dipole cascade model}
\label{ss:mother}
\begin{figure}[t,b]
  \hbox{\vbox{
    \begin{center}
    \mbox{\psfig{figure=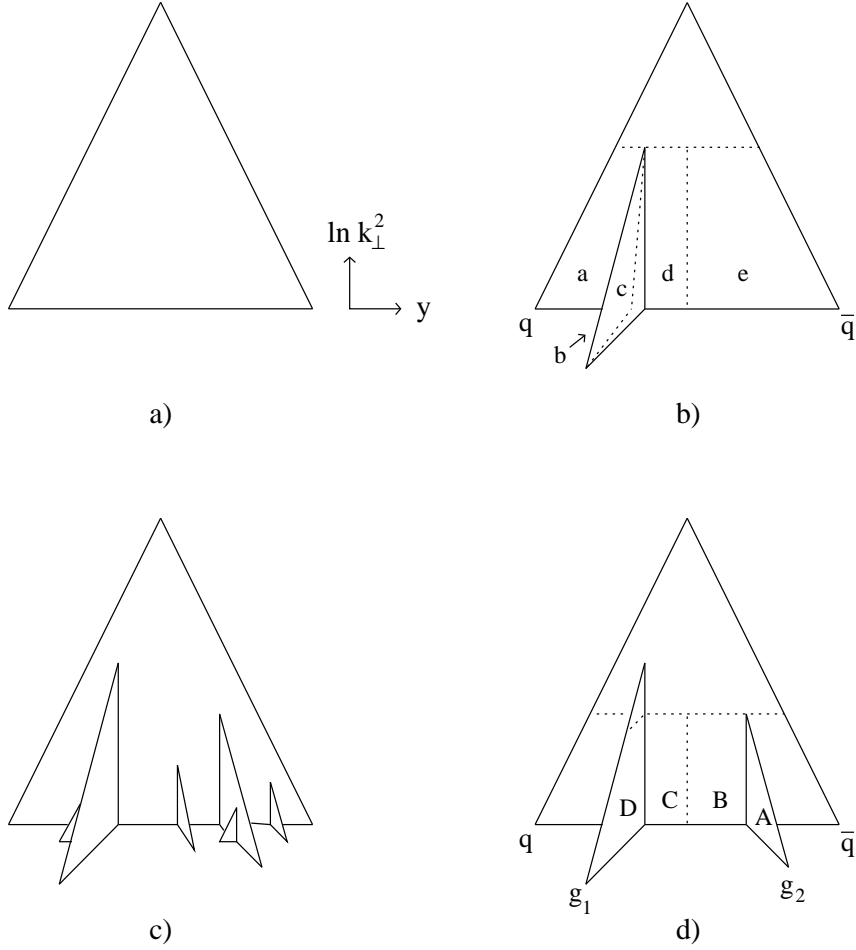,height=12.6cm}}
    \end{center}
  }}
\caption{\em {\em a)} The phase space for gluon emission from a $\epq\epqa$ pair is approximately a triangular region in the $y-\ln k^2_\perp$ plane. {\em b)} If one gluon is emitted the size of the phase space for emission of softer gluons corresponds to this folded surface. {\em c)} Every gluon emission increase the phase space for softer gluons producing a multi-fractal total phase space. {\em d)} Gluon emission associated with surfaces B and C corresponds to emission from an effective quark or an anti-quark charge (triplet or ant-triplet).}
\label{f:phasespace}
\end{figure}

A $\epq\epqa$ pair produced in an \epem-annihilation event emits gluons according to the distribution
\eqbe
dn=C_F\frac{\alpha_s}{2\pi}\frac{x^2_1+x^2_3}{(1-x_1)(1-x_3)}dx_1dx_3
\label{e:emission}
\eqen
where $x_1$ and $x_3$ are the conventional scaled quark and anti-quark momenta. For soft gluons this approximately equals the ``dipole distribution''
\eqbe
dn \approx C_F \frac{\alpha_s}{\pi} d\left[ \ln k^2_\perp \right] dy
\label{e:emissionapprox}
\eqen
where
\eabe
k^2_\perp &=& s(1-x_1)(1-x_3) \nonumber \\
y &=& \frac{1}{2}\ln\frac{1-x_3}{1-x_1}
\eaen
The phase space available is approximately a triangular region, $\left|y\right|<\frac{1}{2}\left[\ln s-\ln k^2_\perp\right]$, in the $y-\ln k^2_\perp$-plane, cf Fig~\ref{f:phasespace}a.

For the emission of two gluons the hardest one is distributed according to Eq~(\ref{e:emission}) or~(\ref{e:emissionapprox}), while the distribution of the softer gluon corresponds to two independent dipoles spanned between the quark and the gluon and between the gluon and the anti-quark respectively. For fixed $k_{\perp2}$ of the second gluon, the available rapidity range in the two dipoles is $\ln\left(s_{\mbox{\scriptsize q} g_1}/k^2_{\perp2}\right)$ and $\ln\left(s_{g_1\overline{\mbox{\scriptsize q}}}/k^2_{\perp2}\right)$, which results in a total rapidity interval
\eqbe
\Delta y =\ln\frac{s_{\mbox{\scriptsize q}g_1}}{k^2_{\perp2}}+\ln\frac{s_{g_1\overline{\mbox{\scriptsize q}}}}{k^2_{\perp2}} = \ln\frac{s}{k^2_{\perp2}} + \ln\frac{k^2_{\perp1}}{k^2_{\perp2}}.
\eqen

Thus the available phase space for the second gluon is increased and corresponds to the folded surface in Fig~\ref{f:phasespace}b. Here the two dipoles correspond to the surfaces a+b and c+d+e respectively. The emission density in the two dipoles is proportional to $C_A/2$, but there is also a correction term with a relative weight $-1/N^2_c$ corresponding to a dipole between the quark and the anti-quark~\cite{ya85}. The result is a density proportional to $C_A/2$ on surfaces b and c and proportional to $C_F=\frac{1}{2}C_A\left[1-1/N^2_c\right]$ on surfaces a, d and e. Thus we see that surfaces b and c correspond to emission from the gluon charges (note that this area is determined by the transverse momentum $k_{\perp1}$ of the first gluon; in the infinite momentum frame this corresponds to the energy times the emission angle) while surfaces a, d and e correspond to emission from a quark (or anti-quark) charge. Here e.g. surfaces a and e can be interpreted as emitted from the quark and the anti-quark. In surface d the quark and the gluon emit coherently as a single triplet charge, due to the soft gluon interference. 

In a similar way the emission of a third gluon corresponds to three dipoles etc. Each emission increases the phase space for softer gluons as leading to a multi-fractal structure illustrated in Fig~\ref{f:phasespace}c~\cite{ba88}. Consider in particular the emission from a $\epq\epqa\ggn_1\ggn_2$ system as shown in Fig~\ref{f:phasespace}d. Here surfaces A, B, C and D correspond to emissions from the dipole between the two gluons. In the same way as above surfaces A and D correspond to emissions from parent gluons, with density $C_A/2$, while in surface C the quark and gluon $\ggn_1$ emit coherently as a triplet (quark) charge with density $C_F$. Similarly in surface B the anti-quark and $\ggn_2$ emit coherently as an anti-triplet charge.

Finally we have to discuss how to associate an emitted gluon (with momentum $k$) to one of the surfaces in Fig~\ref{f:phasespace}d. Assume that the gluon is emitted from the dipole between the gluons $\ggn_1$ and $\ggn_2$. Thus it belongs to one of the surfaces A, B, C, or D. We first separate surface A from B+C+D. We then imagine a dipole between gluon $\ggn_2$ and an effective quark, with momentum $\overline{p}_q+\overline{p}_{\mbox{\scriptsize g}_1} \equiv \overline{P}$. Make a Lorentz boost along $\overline{P}$ such that the momentum $\overline{p}_{\mbox{\scriptsize g}_2}'$ of $\ggn_2$ in the new frame becomes orthogonal to $\overline{P}$. If the momentum $\overline{k}'$ of the new gluon in this frame is closer to $\overline{p}'\!_{\mbox{\scriptsize g}_2}$ than to $\overline{P}$ the new gluon should be associated to surface A, otherwise with surface B+C+D.  (With closer we mean that the relative angle is smaller.) In the same way we can decide whether the new gluon belongs to surface D or to A+B+C. If it belongs to neither A nor D, it obviously belongs to B+C, and thus is emitted by an effective colour triplet charge. This recipe can obviously be used also for a system with an arbitrary number of gluons.

\section{MC Implementation}
The models presented in Sections~\ref{ss:hadron} and~\ref{ss:hardphase} are implemented in a modified version of the Ariadne MC. In the ``cascade reconnection model'' we check after each emission if a reconnection involving the new gluon can give a smaller $\lambda$ measure. If that is the case this configuration will be used with probability 1/8 in the continuation of the cascade. An exception is when two or more gluons originate from a single parent gluon. Such a system must be a colour octet and is therefore not allowed to be separated from the rest of the state. For this reason we keep track of the history of all emissions, as described in Section~\ref{ss:mother}\footnote{A similar reconnection model is developed by L\"onnblad and implemented in Ariadne version 4.07. In this model, however, all emissions from a $\ggn\ggn$ dipole are treated as having a gluon parent. Therefore the reconnection effects become smaller than in our calculations.}.

In the model where reconnection is only allowed after the perturbative phase (the soft reconnection model) there are generally many gluons and many reconnection possibilities. The probability that a certain reconnection should isolate a gluon system originating from a single gluon parent is therefore very small, and has here been neglected in the MC implementation. 

With an implementation of the mother-daughter relation into the MC it is possible to also include the colour suppressed contributions to the emission probabilities. Thus in the MC the emission from an ``effective'' (anti-)quark charge is suppressed by the factor $2C_F/C_A=1-1/N^2_c$. We have, however, not found any observable which is sensitive to this correction. The effect is small, and can generally be compensated by a retuning of the parameters in the fragmentation model. 

\section{Results}
\begin{figure}[t,b]
  \hbox{\vbox{
    \begin{center}
    \mbox{\psfig{figure=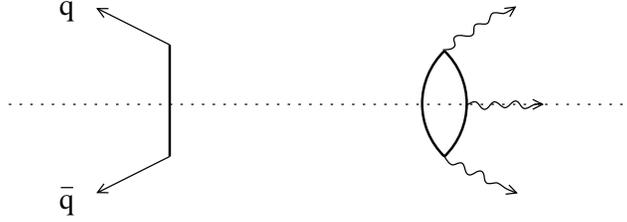,height=2.9cm}}
    \end{center}
  }}
  \caption{\em We are particularly interested in situations where an isolated gluon system moves away from the remaining system, which may result in a central gap in the hadron distribution.}
  \label{f:rapgap}
\end{figure}

The modifications caused by the colour reconnections are generally small and not observable in average event properties. As in~\cite{gg94} we also here try to find a small corner of phase space, where effects could be visible. The events we want to study are those where an isolated gluon system is moving away from the remaining system, as illustrated in Fig~\ref{f:rapgap}.
Such states could reveal themselves by a central gap in the rapidity distribution (cf also ref~\cite{jb92}).

\begin{figure}[t,b]
  \hbox{\vbox{
    \begin{center}
       \mbox{
          \psfig{figure=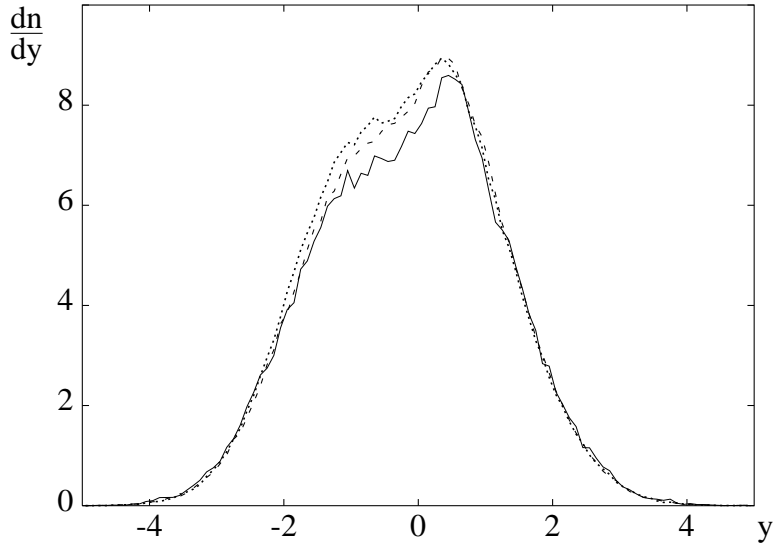,width=10cm}
	}
    \end{center}
  }}
  \caption{\em Rapidity distribution of charged particles, default Ariadne (dotted line), the soft reconnection model (dashed line), and the cascade reconnection model (solid line). Both models are retuned i.e. for average events they reproduce the expected distributions. One million Z$^0$ decay events are analyzed. The rapidity is measured along the direction of the added $\epq \epqa$ momenta.}
  \label{f:rap}
\end{figure}

To know that we have a quark and an anti-quark moving to one side we consider double tagged events with either a $\epb \epba$ or a $\epc\epca$ pair. To obtain a sufficient separation between the two parts in Fig~\ref{f:rapgap} we have studied events where the angle between the quark and the anti-quark is less than 110$^\circ$, and the rapidity axis is chosen along the sum of their momenta. In Fig~\ref{f:rap} we show the rapidity distribution obtained in the different models; normal Ariadne, the soft connection model, and the cascade reconnection model. Here positive $y$-values are in the direction of the $\epq\epqa$ pair. The different Monte Carlo's are tuned to reproduce experimental data.
We see a dip in the distributions from the reconnection models as compared to normal Ariadne. This difference could be a signal of non trivial effects early in the event. 

\begin{figure}[t,b]
  \hbox{\vbox{
    \begin{center}
       \mbox{
          \psfig{figure=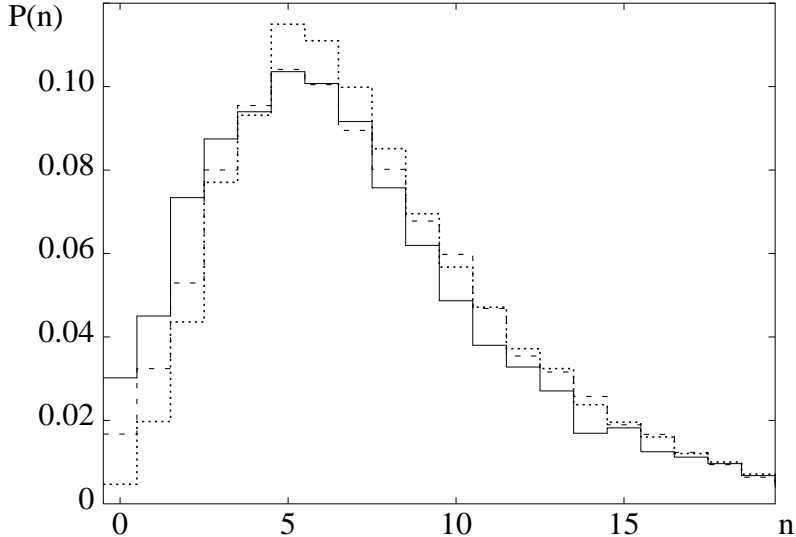,width=10cm}
	}
    \end{center}
  }}
  \caption{\em Charged multiplicity in the rapidity interval [-1,0], default Ariadne (dotted line), the soft reconnection model (dashed line), and the cascade reconnection model (solid line). Both our models are retuned. The distribution is normalized w.r.t. accepted events.}
  \label{f:pny}
\end{figure}

The difference between the models is enhanced if we study, as in ref~\cite{gg94}, the probability to find an event with no charged particles in a central rapidity interval. Choosing the central rapidity region between -1 and 0 we obtain the distributions in Fig~\ref{f:pny},
and we see significant difference between the distributions. It is 7 times more probable to find an event with no particles in the central rapidity bin in the cascade reconnection model as compared to normal Ariadne (in the soft reconnection model this factor becomes 3.5). In Fig~\ref{f:pny} we see that there is also quite a large difference in the probability to find events with only 1 or 2 charged particles in the central rapidity bin.

With the probabilities from Fig~\ref{f:pny} we can calculate the number of events to be expected to have no charged particles in the central rapidity bin. The results for 3 million Z events at LEP1, assuming 10\% double tagging efficiency of the initial heavy quarks, are shown in the table below
\begin{table}[h]
\cebe
\begin{tabular}{|l|ccc|c|}
\hline
	& \multicolumn{3}{c|}{Number of events with $n_{ch}$} & Total number of \\
Model	&\mbox{\hspace{1.8em}0\hspace{1.8em}}&1&\mbox{\hspace{1.8em}2\hspace{1.8em}}& accepted events \\
\hline
Cascade reconnection		& 103 & 154 & 229 & 32700 \\
Soft reconnection		& 62  & 130 & 220 & 37700 \\
Background (normal Ariadne) 	& 17  & 73  & 162 & 37100 \\
\hline
\end{tabular}
\ceen
\caption{\em Expected number of events with 0, 1, or 2 charged particles in a rapidity interval $-1<y<0$. Total events is the number of events which are left after the condition that the angle between the $\epc\epca$ or $\epb\epba$ momenta should be less than 110$^\circ$.}
\end{table}

\section{Conclusions}
We have a very limited knowledge about how the confinement mechanism works in processes like $\epea\epe\ra\epq\epqa\ggn\ldots\ggn$, when there are several identical colour charges. The transition from a Feynman diagram to a string configuration or a cluster chain is not straight forward, as the partons can be connected in several different ways. We do not know if in such a situation Nature chooses a particular configuration at random, or if some configuration is dynamically favoured. The choice of a configuration which differs from the one initially obtained in the generation procedure is often called colour reconnection or recoupling, although the partons have not been connected before, and thus should not be called reconnected.

Also in the perturbative parton cascade we have no well founded recipe for describing the interference effects, which correspond to non-planar diagrams and are caused by identical colour charges. Although these effects are in principle calculable within perturbative QCD, this is at present not possible in practice.

To learn more about how the confinement mechanism and the interference effects work, we have studied two different models. In these models it is assumed that configurations, where the confining string connects partons which are close in phase space, are dynamically favoured.

We are in particular interested in the possibility that a colour singlet gluon system hadronizes isolated from the remainder of the state. A possible signal for such events is a central rapidity gap in events where the angle between the quark and the anti-quark is relatively small. Using double tagged events with heavy $\epc\epca$ or $\epb\epba$ quarks, with a relative angle less than 110$^\circ$, it is possible to find a significant signal, if such events appear in Nature.

If this type of (re)connected states appear in Z decays it may also be an indication that reconnection might appear between the decay products of two W's in the reaction $\epea\epe\ra\gwp\gwm\ra\epq_1\epqa_2\epQ_1\epQa_2$ at LEP2. This would be important e.g. for a precision measurement of the W mass. \vspace{3ex}

\noindent {\bf Acknowledgments} \\
We would like to thank dr L. L\"onnblad and dr T. Sj\"ostrand for valuable discussions.

\end{document}